\documentclass[aps,pra,twocolumn,showpacs,superscriptaddress]{revtex4-1}
\usepackage{amsmath,amssymb,mathrsfs}
\usepackage[bookmarks=false]{hyperref}
\usepackage[dvips]{graphicx}
\usepackage[dvips]{color}
\usepackage{hyperref}
\usepackage{ulem}

\def\be{\begin{equation}}
\def\ee{\end{equation}}
\def\bea{\begin{eqnarray}}
\def\eea{\end{eqnarray}}

\begin{document}
\title{Modulated pair condensate of $p$-orbital ultracold fermions}

\author{Zixu Zhang}
\affiliation{Department of
Physics and Astronomy, University of Pittsburgh, Pittsburgh, PA 15260}

\author{Hsiang-Hsuan Hung}
\affiliation{Department of Physics,
University of California, San Diego, CA 92093}

\author{Chiu Man Ho}
\affiliation{Department of Physics, University of California,
  Berkeley, CA 94720}
\affiliation{Theoretical Physics Group, Lawrence Berkeley National
  Laboratory, Berkeley, CA 94720}
\affiliation{Department of
  Physics and Astronomy, Vanderbilt University, Nashville, TN
  37235}

\author{Erhai Zhao}
\affiliation{Department of
Physics and Astronomy, University of Pittsburgh, Pittsburgh, PA 15260}

\author{W. Vincent Liu}
\affiliation{Department of
Physics and Astronomy, University of Pittsburgh, Pittsburgh, PA 15260}

\date{\today}

\begin{abstract}
We show that an interesting of pairing occurs for
spin-imbalanced Fermi gases under a specific experimental condition---the
spin up and spin down Fermi levels lying within the $p_x$ and $s$
orbital bands of an optical lattice, respectively. The
pairs condense at a finite momentum equal to the sum of the two Fermi momenta of spin up and spin down fermions and form a $p$-orbital pair condensate.
This $2k_F$ momentum dependence
has been seen before in the spin- and charge- density waves, but it differs from the
usual $p$-wave superfluids such as $^3$He, where the orbital symmetry
refers to the relative motion within each pair. Our conclusion is based on the density matrix renormalization group analysis for the one-dimensional (1D) system and mean-field theory for the quasi-1D system. The phase diagram of the quasi-1D system is calculated, showing that the $p$-orbital pair condensate occurs in a wide range of fillings.  In the strongly attractive limit, the system realizes an unconventional BEC beyond Feynman's no-node theorem. The possible experimental signatures of this phase in molecule projection experiment are discussed.
\end{abstract}

\pacs{03.75.Ss, 71.10.Fd, 37.10.Jk, 05.30.Fk}

\maketitle
\section{introduction}
Pairing with mismatched Fermi surfaces has long fascinated researchers
in the fields of heavy fermion and organic superconductors, color
superconductivity in quark matter \cite{RevModPhys.76.263}, and, most
recently, ultracold Fermi gases with spin imbalance
\cite{giorgini:1215,Sheehy20071790,0034-4885-73-7-076501,PhysRevLett.99.250403}.  In a classic
two-component model for superconductivity, the mismatch arises from
the spin polarization of fermions in the same energy band. Its
effect was predicted to produce intriguing, unconventional superfluids
such as the Fulde-Ferrell-Larkin-Ovchinnikov (FFLO) \cite{PhysRev.135.A550, fflo2}, deformed Fermi
surface \cite{PhysRevLett.88.252503, PhysRevA.72.013613}, and breached pair phases \cite{PhysRevLett.90.047002, PhysRevLett.94.017001}. The limiting case of large spin imbalance was also studied to explore
the formation of Fermi polarons \cite{PhysRevLett.102.230402}.  In
parallel, the behavior of particles in the higher orbital bands
of optical lattices, due to large filling factors, thermal
excitations or strong interactions, is widely studied for novel
orbital orderings of both bosons \cite{liu:013607,PhysRevA.72.053604,
kuklov:110405} and fermions \cite{zhao:160403,wu:200406} with
repulsive interactions. Recently, interband pairing of unpolarized
fermions was shown theoretically to give rise to Cooper pair density
waves \cite{PhysRevB.81.012504}.

In this article, we report a fermion pairing
phase resulting from the interplay of Fermi surface mismatch and
$p$-orbital band physics. In such a phase, the pair condensate
wave function is spatially modulated and has a $p$-wave symmetry.
This phase arises in an attractive two-component Fermi
gas on anisotropic optical lattices under a previously unexplored condition of spin imbalance.  Namely the majority ($\uparrow$) spin and the
minority ($\downarrow$) spin occupy up to Fermi levels lying in the
$p_x$ and $s$ bands, respectively.
We show that pairings take place near the respective Fermi surfaces of
the spin $\uparrow$ fermions in $p_x$ band and $\downarrow$ fermions in $s$ band.
This induces a modulated $p$-orbital pair condensate that differs from the usual $p$-wave superfluids such as $^3$He.
The state requires only an on-site isotropic contact interaction and the pair is a spin singlet, while the $^3$He $p$-wave superconductivity
has to involve anisotropic interaction and spin triplet.  The
modulation wave vector of the order parameter is ${Q}\approx k_{F\uparrow}+k_{F\downarrow}$, where $k_{F\uparrow}$, $k_{F\downarrow}$ are Fermi momenta for spin $\uparrow$ and $\downarrow$ species, respectively. This $2k_F$ momentum dependence
is an unprecedented signature in superfluids other than the spin- and charge- density waves.
In the strongly attractive limit,  tightly bounded pairs condense
at finite momentum $Q$, which realizes an unconventional Bose-Einstein condensate beyond
Feynman's no-node theorem \cite{liu:013607, PhysRevA.72.053604, congjun:modphyslett, kuklov:110405,
radzihovsky:095302}.

\section{model}
The system under consideration is at zero temperature and consists of two-component fermions in a
three-dimensional (3D) cubic optical lattice with lattice constant $a$, described by the
Hamiltonian
\begin{eqnarray}
H&=&\sum_{\sigma} \int
d^3\mathbf{x}\psi^{\dag}_{\sigma}(\mathbf{x}) [-\frac{\hbar^2}{2m}\nabla^2+V(\mathbf{x})-\mu_{\sigma}] \psi_{\sigma}(\mathbf{x})\nonumber\\
&&+g\int
d^3\mathbf{x}\psi_{\uparrow}^{\dag}(\mathbf{x}) \psi_{\downarrow}^{\dag}(\mathbf{x})\psi_{\downarrow} (\mathbf{x})\psi_{\uparrow}(\mathbf{x}).
\label{eq:generalham}
\end{eqnarray}
Here $\psi_{\sigma}(\mathbf{x})$ is the fermionic field operator at
$\mathbf{x}$ with spin $\sigma=\uparrow, \downarrow$, $V(\mathbf{x})$
is the lattice potential, $\mu_{\sigma}$ is the chemical potential for
spin $\sigma$ fermions, and $g<0$ is the contact attraction which can be tuned by the Feshbach resonance. In
particular, we consider the case where the lattice potential in the $x$
(parallel) direction is much weaker than the other two (transverse)
directions, so the system behaves quasi-one-dimensionally.

We expand
$\psi_{\sigma}(\mathbf{x})=
\sum_{n\mathbf{r}}\phi_n(\mathbf{x}-\mathbf{r})c_{n\mathbf{r}}$,
where $\phi_n(\mathbf{x}-\mathbf{r})$ is the $n$th band Wannier
function at lattice site $\mathbf{r}$ with $c_{n\mathbf{r}}$ the
annihilation operator in Wannier basis. As a result, we obtain the
usual attractive Hubbard model with nearest-neighbor hopping between $i$th site with orbital band $\alpha$ and $j$th site with orbital band $\beta$
\begin{equation}
t_{\alpha \beta}=-\int d^3 \mathbf{x} \phi_{\alpha}^*(\mathbf{x}-\mathbf{r}_i) \left[-\frac{\hbar ^2 \nabla ^2}{2m} + V(\mathbf{x})\right]\phi_{\beta}(\mathbf{x}-\mathbf{r}_j)
\label{eq:hopping}
\end{equation}
and
on-site attraction between orbitals
\begin{equation}
U_{\alpha \beta \gamma \eta}=g\int d^3\mathbf{x}\phi_{\alpha}^*(\mathbf{x}-\mathbf{r}_i) \phi_{\beta}^*(\mathbf{x}-\mathbf{r}_i) \phi_{\gamma}(\mathbf{x}-\mathbf{r}_i) \phi_{\eta}(\mathbf{x}-\mathbf{r}_i).
\label{eq:onsiteint}
\end{equation}

The lowest two energy bands are the $s$ and
$p_x$ band (the $p_y$ and $p_z$ band are much higher in energy because
of  tighter confinement in the transverse directions). For brevity
the $p_x$ band is simply called $p$ band in the
following. By filling fermions with spin $\uparrow$ to the $p$ band and spin $\downarrow$ to the $s$ band, the
Hamiltonian becomes
\begin{eqnarray}
H_{sp}&=&-\sum_{\langle\mathbf{r,r'}\rangle} (t_s^{\parallel}
S_{\mathbf{r}}^{\dag}S_{\mathbf{r'}}-t_p^{\parallel} P_{\mathbf{r}}^{\dag}
P_{\mathbf{r'}}+h.c.)-\mu_s\sum_{\mathbf{r}} n_{\mathbf{r}}^s\nonumber\\
&&-\sum_{\langle\mathbf{r,r''}\rangle} (t_{s}^{\perp}
S_{\mathbf{r}}^{\dag}S_{\mathbf{r''}}+t_p^{\perp} P_{\mathbf{r}}^{\dag}
P_{\mathbf{r''}}+h.c.)-\mu_p\sum_{\mathbf{r}} n_{\mathbf{r}}^p\nonumber\\
&&+\omega_b\sum_{\mathbf{r}} n_{\mathbf{r}}^p+U_{sp}\sum_{\mathbf{r}} n_{\mathbf{r}}^s
n_{\mathbf{r}}^p.
\label{eq:ham}
\end{eqnarray}
Here, $\langle\mathbf{r,r'}\rangle$ and $\langle\mathbf{r,r''}\rangle$
denote the nearest neighboring lattice sites in parallel and
transverse directions. $t_s^{\parallel}$ and $t_p^{\parallel}$ are the
hopping amplitudes along the parallel direction for the $s$- and
$p$-band fermions respectively, while
$t_s^{\perp}=t_p^{\perp}=t^{\perp}$ are the hopping amplitudes in
transverse directions. $S_{\mathbf{r}}$ ($P_{\mathbf{r}}$) is the
annihilation operator at lattice site $\mathbf{r}$ for $s$-band
$\downarrow$ ($p$-band $\uparrow$)
fermions. $n_{\mathbf{r}}^s=S_{\mathbf{r}}^{\dag}S_{\mathbf{r}},
n_{\mathbf{r}}^p=P_{\mathbf{r}}^{\dag}P_{\mathbf{r}}$ are the number
operators, and $\mu_s,\mu_p$ are the corresponding chemical
potentials. $U_{sp}$ is the attractive on-site interaction between $s$-
and $p$-band fermions and can be tuned by changing the scattering
length using Feshbach resonance. $\omega_b$ is related to the band
gap. In the tight binding region we assume $\omega_b\gg |U_{sp}|$, and
consequently the $s$-band fully filled spin $\uparrow$ fermions are
dynamically inert and not included in the $H_{sp}$.

\section{DMRG calculation for 1D case}
\begin{figure}[t]
\includegraphics[width=193pt,height=113pt]{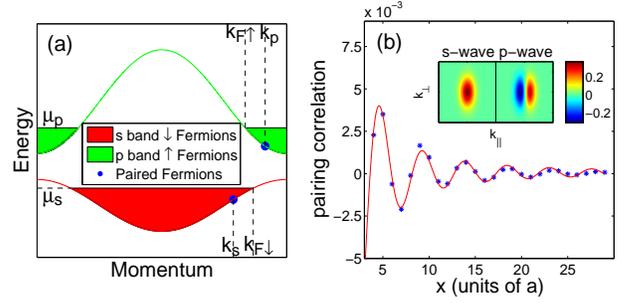}
\caption{(Color online) (a) A schematic illustration showing the pairing between $s$- and $p$-band
fermions. The $s$ band is also fully occupied with $\uparrow$ fermions (not shown). (b) The spatial
variation of the pairing correlation $C(x)$ for $N_s=49$, $N_p=15$ according to DMRG. The blue
scatters are the DMRG result and the solid line is the fitting using function
$a\cos(qx+b)/x^{\eta}+c$. The inset in (b) shows  the $s$- and $p$-wave Wannier functions in momentum
space, which are elongated in the transverse direction (in real space they are compressed in
transverse direction). The $s$-wave Wannier function has even parity while the $p$-wave Wannier function has odd parity.}
\label{fig:cartoon}
\end{figure}

First we consider the pairing problem in the simplest case of 1D
($t^{\perp}=0$), which is schematically shown in
Fig.~\ref{fig:cartoon}(a). The two relevant Fermi momenta are
$k_{F\downarrow}$ (for $s$-band $\downarrow$ fermions) and
$k_{F\uparrow}$ (for $p$-band $\uparrow$ fermions). From a weak
coupling point of view, to pair fermions of opposite spin near their
respective Fermi surfaces, the Cooper pairs have to carry finite
center-of-mass momentum (CMM) due to Fermi surface
mismatch. Furthermore, in order for all Cooper pairs to have roughly
the same CMM, the only choice is to pair fermions of opposite
chirality. Note that the dispersion of $p$ band is inverted with
respect to the $s$ band, so pairing occurs between fermions with
momenta of the same sign but opposite group velocities.  These
elementary considerations show that the CMM of the pair should be approximately the
sum of two Fermi momenta,
\begin{equation}
Q\approx k_{F\uparrow}+k_{F\downarrow}\,.
\label{eq:cartooncouple}
\end{equation}
This result differs from that of the usual
one-dimensional spin imbalanced fermions within the same
band, where the FFLO pair
momentum is the difference,
${Q}\approx|k_{F\uparrow}-k_{F\downarrow}|$, as found in
a two-leg-ladder system \cite{feiguin:076403}.

Mean-field theory and weak coupling consideration can provide only a
qualitative picture for 1D problems. To unambiguously identify the
nature of the ground state, we use density matrix renormalization group (DMRG) to compute the pair
correlation function. In the numerical calculations, we used
parameters $t_s^{\parallel}=1$ as the unit of energy, $t_p^{\parallel}=8$, $\mu_s=1.7$,
$\mu_p-\omega_b=-11$, in which the ratio between $t_s$ and $t_p$ is chosen according to typical tight-binding bandwidth ratio. $U_{sp}$ is tunable with Feshbach resonance and in the following calculation we will focus on $U_{sp}=-9$ \cite{footnote}.
The truncation error is controlled in the order of
$10^{-7}$ or less. Equation~\eqref{eq:cartooncouple} predicts
$Q\approx k_{F\uparrow}+k_{F \downarrow}=0.435
\pi/a$. Figure~\ref{fig:cartoon}(b) shows the pairing correlation
function in real space $C_{ij}=\langle S^{\dag}_i P^{\dag}_i P_j
S_j\rangle$ as a function of $x=|i-j|$ for a chain of $N=60$ sites
with open boundary condition, where the indices $i$ and $j$ are real
space positions. Since the system only has algebraic order, $C(x)$
decays with $x$ according to a power law. On top of this, however,
there is also an obvious oscillation. A curve fit with formula
$C(x)=a\cos(qx+b)/x^{\eta}+c$, shown in Fig.~\ref{fig:cartoon}(b),
yields a period of $q=0.438 \pi/a$, which is in good agreement with
the wave number given by Eq.~\eqref{eq:cartooncouple} before. The
Fourier transform of the pair correlation function
\be
C_q={1\over N}\sum_{i,j} e^{iq(i-j)} C_{ij}
\label{eq:moleculedmft}
\ee
is peaked at $q=0.426\pi/a$ (to be plotted in Sec.~\ref{sec:molecule}). These
features of the pair correlation function are the signature of the existence of the $2k_F$ CMM pairing in our system  \cite{feiguin:220508,zhao:063605}.

\section{Meanfield analysis for quasi 1d case}
Now we move on to the quasi-1D system where a weak transverse hopping $t^\perp\ll t^\parallel$ is
added. We carry out a mean-field analysis of Hamiltonian $H_{sp}$ by introducing the $s$-$p$ pairing
order parameter
\begin{equation}
\Delta_{\mathbf{r}}=U_{sp}\langle S_{\mathbf{r}} P_{\mathbf{r}}\rangle,
\label{eq:selfcons}
\end{equation}
where $\langle...\rangle$ means the ground-state expectation value.
Two different trial ground states are investigated, the exponential wave $\Delta_{\mathbf{r}}=\Delta e^{i\mathbf{Q\cdot r}}$, which is analogous to the Fulde-Ferrell phase and the cosine wave $\Delta_{\mathbf{r}}=\Delta \cos \mathbf{Q\cdot r}$, which is analogous to the Larkin-Ovchinnikov phase. $\mathbf{Q}$ and $\Delta$ are determined
self-consistently by minimization of ground-state free energy $\langle H_{sp}\rangle$. Transverse
hopping introduces a small Fermi surface curvature and spoils the perfect nesting condition as in the
pure 1D problem above. However, the curvature is small for weak $t^\perp$. Thus, we expect
$\mathbf{Q}$ pointing almost along the parallel direction, $\mathbf{Q}=Q(1,0,0)$, in order to
maximize the phase space of pairing.

The mean-field Hamiltonian for the exponential wave can be diagonalized in momentum space by standard
procedure. We get the ground state energy
\begin{eqnarray}
\langle
H_{sp}\rangle=\sum_{\mathbf{k},\gamma=\pm}\Theta (-\lambda_{\mathbf{k}}^{(\gamma)})\lambda_{\mathbf{k}}^{(\gamma)} +\sum_{\mathbf{k}}\xi^p_{\mathbf{k}}-\frac{N^3\Delta^2}{U_{sp}}
\label{eq:exp}
\end{eqnarray}
with the self-consistent gap equation for $\Delta$
\begin{eqnarray}
1=\frac{
U_{sp}}{N^3}\sum_{\mathbf{k}}\frac{\Theta (-\lambda_{\mathbf{k}}^{(+)})-\Theta(-\lambda_{\mathbf{k}}^{(-)})} {\sqrt{4\Delta^2+(\xi_{\mathbf{k}}^s+\xi_{\mathbf{Q-k}}^p)^2}}.
\label{eq:gapexp}
\end{eqnarray}
Here, $\mathbf{k}$ is lattice momentum, $N^3$ is the total number of sites, $\Theta$ is a step
function, and $\lambda_{\mathbf{k}}^{(\pm)}=\frac{1}{2}[\xi_{\mathbf{k}}^s- \xi_{\mathbf{Q-k}}^p \pm
\sqrt{4\Delta^2+(\xi_{\mathbf{k}}^s+\xi_{\mathbf{Q-k}}^p)^2}]$
 is the eigenenergy of the Bogoliubov quasiparticles. As evident from these formulas, the pairing
 occurs between an $s$-band fermion of momentum $\mathbf{k}$ and a $p$-band fermion of momentum
 $\mathbf{Q-k}$ with dispersion $\xi^s_{\mathbf{k}}=-2t_s^{\parallel}\cos k_x a-2t^{\perp}\cos k_y
 a-2t^{\perp}\cos k_z a-\mu_s$ and $\xi^p_{\mathbf{k}}=2t_p^{\parallel}\cos k_x a-2t^{\perp}\cos k_y
 a-2t^{\perp}\cos k_z a-\mu_p+\omega_b$, respectively.

The cosine wave is spatially inhomogeneous. A full mean-field
analysis requires solving the Bogoliubov-de Gennes equation to
determine the gap profile self-consistently. Here we are interested only in computing the free energy for the ansatz
$\Delta_{\mathbf{r}}=\Delta \cos \mathbf{Q\cdot r}$ to compare with
the exponential wave case. Thus, it is sufficient to numerically diagonalize the
full Hamiltonian Eq.~\eqref{eq:ham} for a finite size lattice. We introduce
a vector of dimension $2N$
\begin{equation}
\alpha^{\dag}_{k_yk_z}=(S^{\dag}_{k_x^1
k_yk_z}...S^{\dag}_{k_x^Nk_yk_z},P_{k_x^1,-k_y,-k_z}...P_{k_x^N,-k_y,-k_z}),
\label{eq:alpha}
\end{equation}
where $k_x^n=2\pi n/Na$ is the discrete momentum in the $x$ direction.
The components of $\alpha$ obey anticommutation relation
$\{\alpha_{k_yk_z}^{\dag(m_1)},\alpha_{k_yk_z}^{(m_2)}\}=\delta_{m_1m_2}$,
where $m_1,m_2$ labels the corresponding operator component of
$\alpha$. The Hamiltonian takes the compact form
$H_{sp}=\sum_{k_yk_z}\alpha^{\dag}_{k_yk_z}\mathcal{H}_{k_yk_z} \alpha_{k_yk_z}+\sum_{\mathbf{k}}\xi^p_{\mathbf{k}}- (1+\delta_{Q,-Q})N^3\Delta^2/2U_{sp}$.
Since $\mathcal{H}_{k_yk_z}$ is real and symmetric, it can be
diagonalized by an orthogonal transformation
$\alpha_{k_yk_z}=\mathcal{D}_{k_yk_z}\beta_{k_yk_z}$ to yield $2N$
eigenvalues $E_{k_yk_z}^l$. The new operators $\beta_{k_yk_z}$
automatically obey the fermionic anticommutation relationship
$\{\beta_{k_yk_z}^{\dag(m_1)},\beta_{k_yk_z}^{(m_2)}\}=\delta_{m_1m_2}$.
We get the ground state energy,
\begin{eqnarray}
\langle
H_{sp}\rangle&=&\sum_{k_y,k_z}\sum_{l=1}^{2N}E_{k_yk_z}^l \Theta(-E_{k_yk_z}^l)+\sum_{\mathbf{k}}\xi^p_{\mathbf{k}}\nonumber\\
&&-\frac{N^3\Delta^2}{2U_{sp}}(1+\delta_{-Q,Q})\,,
\label{eq:cos}
\end{eqnarray}
and the gap equation,
\begin{equation}
\Delta=\frac{2U_{sp}}{N^3(1+\delta_{-Q,Q})}\sum_{\mathbf{k}} \sum_l\mathcal{D}_{k_yk_z}^{m_1,l}\mathcal{D}_{k_yk_z}^{m_1',l} \Theta(E_{k_yk_z}^l).
\label{eq:gapcos}
\end{equation}
Here, $l$ labels the eigenenergy, and $m_1$, $m_1'$ labels the matrix elements corresponding to the
original $S$, $P$ operators in the gap equation.

\begin{figure}[t]
\includegraphics[width=240pt,height=185pt]{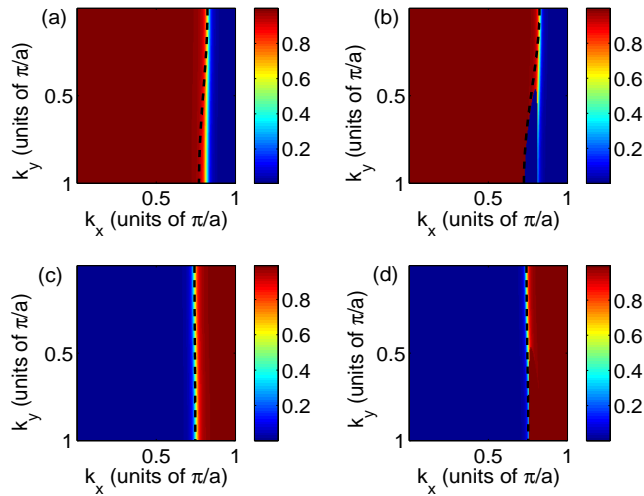}
\caption{(Color online) The occupation of $s$ and $p$ band within the paired state for different transverse hopping $t_\perp$.
Only the first quadrant of the Brillouin zone in the $k_x-k_y$ plane is shown,
$k_z=\pi/a$. The black dashed lines indicate the ``bare" Fermi
surfaces for corresponding noninteracting
fermions  ($U_{sp}=0$). (a) $\langle S_{\mathbf{k}}^{\dag}S_{\mathbf{k}} \rangle$
for $t^{\perp}=0.05$; (b) $\langle S_{\mathbf{k}}^{\dag}S_{\mathbf{k}}
\rangle$ for $t^{\perp}=0.1$; (c) $\langle
P_{\mathbf{k}}^{\dag}P_{\mathbf{k}} \rangle$ for $t^{\perp}=0.05$; (d)
$\langle P_{\mathbf{k}}^{\dag}P_{\mathbf{k}} \rangle$ for
$t^{\perp}=0.1$.}
\label{fig:fs}
\end{figure}

The parameters used in the mean-field calculations are the same as in the 1D
case with small $t^{\perp}$'s added, and we still expect that the order parameter
has the momentum around $0.435\pi/a$ as before. By self-consistently solving for $Q$ and $\Delta$, in the case $t^{\perp}=0.05$, the ground state is the cosine wave phase with $Q=0.433\pi/a$ and $\Delta=0.822$. The ground state energy per site is $-2.5927$, lower than the noninteracting value $-2.5896$. When $t^{\perp}=0.1$, the ground state is also the cosine wave phase with $Q=0.433 \pi/a$ and $\Delta=0.542$. The ground state energy per site is $-2.5955$, lower than the noninteracting value $-2.5949$. These results confirm that (i) the
cosine wave state has lower energy than the exponential wave state, (ii) the order parameter
has the momentum close to the prediction of
Eq.~\eqref{eq:cartooncouple}, and (iii) larger transverse hopping tends to
destroy the $p$-orbital pair condensate since the energy gain for larger transverse hopping is much smaller than for smaller transverse hopping.

An interesting feature of the $p$-orbital pair condensate in quasi-1D is the
possible existence of Fermi surfaces with gapless energy
spectrum. We monitor the fermion occupation
number, i.e. $\langle S_{\mathbf{k}}^{\dag}S_{\mathbf{k}} \rangle$ and
$\langle P_{\mathbf{k}}^{\dag}P_{\mathbf{k}}\rangle$ for increasing
transverse hopping. The results are shown in Fig.~\ref{fig:fs}. For
small $t^{\perp}$, they take the usual BCS form and vary smoothly from 1 (red) to 0 (blue)
across the bare Fermi surface (with interaction turned off), as shown
in Figs.~\ref{fig:fs}(a) and \ref{fig:fs}(c) for $t^\perp=0.05$. For larger
transverse hopping, sharp Fermi surfaces characterized by a sudden
jump in $\langle S_{\mathbf{k}}^{\dag}S_{\mathbf{k}} \rangle$ and
$\langle P_{\mathbf{k}}^{\dag}P_{\mathbf{k}}\rangle$ appear. This is
clearly shown in Figs.~\ref{fig:fs}(b) and \ref{fig:fs}(d) for $t^\perp=0.1$ as the occupation number changes discontinuously from 1 (red) to 0 (blue). It can
be understood qualitatively as follows. As $t^{\perp}$ increases, the
original Fermi surfaces acquire a larger curvature in the transverse
directions and the pairing condition in Eq.~\eqref{eq:cartooncouple}
cannot be satisfied everywhere anymore. Therefore in some regions
fermions are not paired and Fermi surfaces survive. One should
also note that the calculation is based on the assumption that
$t^{\perp} \ll t^{\parallel}$, which predicts that $\mathbf{Q}$ is in
the parallel
direction. This prediction should fail as $t^{\perp}$ increases beyond
certain critical values.

\section{Phase Diagram}
Now, we systematically explore the phases of our system for general band filling and spin imbalance. Since we have $s$- and $p$- bands with different bandwidths, we introduce
two dimensionless quantities for the chemical potentials $\mu_s$ and $\mu_p$
\begin{eqnarray}
\tilde{\mu}_s&=&\frac{\mu_s}{2t_s}=\frac{\mu_s}{2},\nonumber\\
\tilde{\mu}_p&=&\frac{\mu_p-\omega_b}{2t_p}=\frac{\mu_p-\omega_b}{16}.
\label{eq:renormu}
\end{eqnarray}
Thus, for a non-interacting system, $-1 < \tilde{\mu}_s, \tilde{\mu}_p <1$ control the filling for the $s$ and $p$-band fermions
respectively. We then define the quantities
\begin{eqnarray}
\mu&=&\frac{\tilde{\mu}_s+\tilde{\mu}_p}{2},\nonumber\\
h&=&\frac{\tilde{\mu}_s-\tilde{\mu}_p}{2},
\label{eq:muandh}
\end{eqnarray}
as the parameters controlling the average filling and polarization in the phase diagram. The phase at $-\mu, -h$ is the same as the state at $\mu, h$, since the transformation $\mu,h \rightarrow -\mu, -h$ gives $\mu_s, \mu_p \rightarrow -\mu_s, -\mu_p$, and the mean-field Hamiltonian with $\mu_s, \mu_p$ is identical to Hamiltonian with $-\mu_s, -\mu_p$ via a particle-hole transformation up to a constant.

\begin{figure}[t]
\includegraphics[width=215pt,height=80pt]{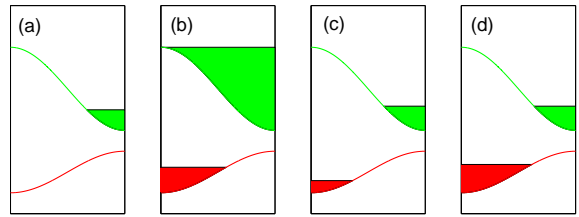}
\caption{(Color online) Band occupation for the four possible phases in the system. The band colored in red represents the $s$ band occupied by spin $\downarrow$ fermions and the band colored in green represents the $p$ band occupied by $\uparrow$ fermions. The spin $\uparrow$ fermions in the $s$ band are not shown since they are inert. (a) Normal phase I (N1) with one band empty and the other partially filled. (b) Normal phase II (N2) with one band fully filled and the other partially filled.  (c)  Commensurate $p$-orbital pair condensate (CpPC) with both bands partially filled. The occupation numbers are the same.  (d) Incommensurate $p$-orbital pair
condensate (IpPC) with both bands partially filled. The occupation numbers are different.}
\label{fig:phases}
\end{figure}

We have four possible phases in such a system as shown in Fig.~\ref{fig:phases}. As before, we ignored the inert fully filled $s$ band of spin $\uparrow$ fermions. We consider the $p$ band of spin $\uparrow$ fermions and $s$ band of spin $\downarrow$ fermions. When one of these two bands is empty and the other is filled, the pairing does not happen and we call it normal phase I (N1) as in Fig.~\ref{fig:phases}(a). When one of these two bands is fully filled and the other is partially filled, the pairing also does not happen since the fully filled band is inert. We call it normal phase II (N2) as in Fig.~\ref{fig:phases}(b). When both of them are partially filled, fermions near Fermi surfaces from the two bands will be paired and the system is in superfluid phases as shown in Figs.~\ref{fig:phases}(c) and \ref{fig:phases}(d). In the superfluid regime, when $h$ is small, the pairing momentum prefers $Q=\pi/a$ and we call it commensurate $p$-orbital pair condensate (CpPC). It is a special case of the $p$-orbital pair condensate, where the occupation numbers of $s$-band spin $\downarrow$ fermions and $p$-band spin $\uparrow$ fermions are the same. It is similar to the conventional unpolarized pairing (BCS), where the spin $\uparrow$ fermions and spin $\downarrow$ fermions have the same population. However, in BCS pairing the CMM of the pair has the property $Q=0$, while here $Q=\pi/a$. To understand the momentum $\pi/a$ preference, note that in conventional BCS case, the two species of fermions have the same energy spectrum and the pairing is between two fermions with opposite momenta, which leads to the CMM of pair $Q=0$. Here, the structure of energy spectrum of $p$ band is different from $s$ band. The equal occupation numbers mean $k_{F\uparrow}=\pi/a-k_{F\downarrow}$, which gives rise to $Q=k_{F\uparrow}+k_{F\downarrow}=\pi/a$, as shown in Fig.~\ref{fig:phases}(c). At last, when $h$ is large, the pairing momentum stays at a general $Q\approx k_{F\uparrow}+k_{F\downarrow}$ and the occupation number for the two species of fermions differ. We call it incommensurate $p$-orbital pair condensate (IpPC) as shown in Fig.~\ref{fig:phases}(d).

To determine the phases, we minimize the free energy as a function of the pairing amplitude $\Delta$ and pairing momentum $Q$ by mean-field analysis using the cosine wave function as outlined in the previous section. When the minimum is realized at $\Delta=0$, it is normal phase. When $\Delta$ is finite, there are two possibilities. When $Q=\pi/a$, it is CpPC. When $Q\neq\pi/a$, it is IpPC. For the transition between superfluid and normal phase, and the transition between CpPC and IpPC, the behaviors of free energy show that the phase transitions are first order in a lattice system. Between the superfluid and normal phases, near the phase transition, $\Delta$ changes suddenly from $0$ to finite, and the free energy shows two local minima at $\Delta=0$ and $\Delta \neq 0$. Between CpPC and IpPC, the pairing momentum changes from $Q=\pi/a$ to $Q\neq\pi/a$ discontinuously, and the free energy as a function of $Q$ also has two local minima at $Q=\pi/a$ and $Q \neq \pi/a$. Thus, they are first-order phase transitions according to our mean field analysis. Therefore, we can determine the phase boundaries between normal phase and superfluid phase by monitoring $\Delta$ changing from zero to finite. We can also monitor $Q$ changing from $Q=\pi/a$ to $Q\neq \pi/a$ to determine the phase boundaries between CpPC and IpPC.

\begin{figure}[t]
\includegraphics[width=172pt,height=150pt]{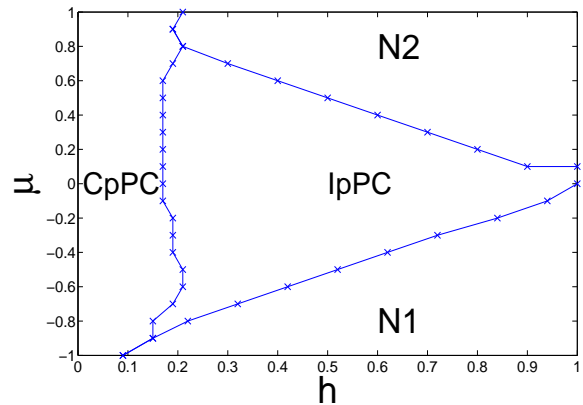}
\caption{(Color online) The phase diagram of the $p$-orbital pair condensate for $t^{\perp}=0.05$. $\mu$ and $h$ are defined in the main text. CpPC: the $s$ band of spin $\downarrow$ fermions and the $p$ band of spin $\uparrow$ fermions have the same occupation numbers. IpPC: the $s$ band of spin $\downarrow$ fermions and the $p$ band of spin $\uparrow$ fermions have different occupation numbers. N1 with the $p$ band of spin $\uparrow$ fermions empty and the $s$ band of spin $\downarrow$ fermions partially filled. N2 with the $p$ band of spin $\uparrow$ fermions partially filled and the $s$ band of spin $\downarrow$ fermions fully filled.}
\label{fig:pd}
\end{figure}

In Fig.~\ref{fig:pd}, we present a phase diagram for $t^{\perp}=0.05$. The x's in Fig.~\ref{fig:pd} show the data points for the phase boundary obtained from the numerical procedure, and by connecting them we get the phase boundaries. An illustrative physical understanding about this phase diagram is as follows. In Fig.~\ref{fig:pd}, when chemical potential difference $h$ is small and the two bands are still partially filled to ensure the pairing, the system tends to stay in CpPC where $Q=\pi/a$. It is similar to the conventional BCS superfluid case. As $h$ becomes larger, as long as the average filling $\mu$ is not too large or small and the two bands are still both partially filled, the pairing persists despite the spin imbalance and the system is in IpPC. If $\mu$ gets more and more negative, the average filling becomes smaller and smaller, and at certain $\mu, h$, $p$ band of spin $\uparrow$ fermions will be empty and the system will become N1 without pairing. Similarly, when $\mu$ is large and positive, the average filling is very high and at certain $\mu, h$, the $s$ band of spin $\downarrow$ fermions will be fully occupied, and the system becomes N2 without pairing. The almost straight phase boundaries in Fig.~\ref{fig:pd} between IpPC and normal phases indicate that these phase transitions are due to the change of band occupation as empty $\leftrightarrow$ partially filled $\leftrightarrow$ fully filled. In Fig.~\ref{fig:pd}, the phase boundary between IpPC and N1 corresponds to the critical condition that the $s$ band of spin $\downarrow$ fermions is partially filled while the $p$ band of spin $\uparrow$ fermion becomes empty, and the almost straight phase boundary corresponds to the condition that $\tilde{\mu}_p=\mu-h=-1$ (but, as before, this is only an approximate argument due to the presence of interaction). Similarly, the almost straight phase boundary between IpPC and N2 corresponds to the condition that the $s$ band of spin $\downarrow$ fermions becomes fully filled, while the $p$ band of spin $\uparrow$ fermions is partially filled, or $\tilde{\mu}_s=\mu+h=1$. All the phase transition lines in Fig.~\ref{fig:pd} are mean field results, and these straight lines are expected to be corrected by quantum critical fluctuations. The phase diagram shows that the $p$-orbital pair condensate happens in large parameter regimes and is closely related to the band and orbital properties in the optical lattice systems.

\section{Signature of the $p$-orbital pair condensate in Molecule Projection Experiment}
\label{sec:molecule}
\begin{figure}[t]
\includegraphics[width=226pt,height=102pt]{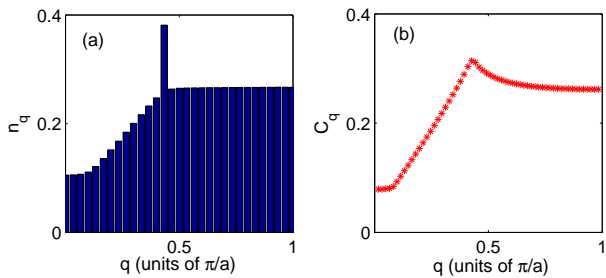}
\caption{(Color online)
(a) The momentum distribution function $n_q$  of projected molecules for a quasi-1D system with
$t^{\perp}=0.05$ (all other parameters are same as before) according to mean field theory. Here,
$q=q_x$, $q_y=q_z=0$.
(b) Pair correlation function $C_q$ for a 1D chain of $N=60$ sites obtained by DMRG.
The peak is located at $0.433 \pi/a$ in both figures, which corresponds to the value $k_{F\uparrow}+k_{F\downarrow}=(N_s+N-N_p)\pi/Na$ for $N_s=49$ and $N_p=15$.
}
\label{fig:pair}
\end{figure}

The $p$-orbital pair condensate phase can inspire
important experimental signatures for finite momentum condensation
of bosonic molecules in higher orbital bands. By
fast sweeping the magnetic field (and thus the interaction) from the
BCS region to the deep BEC region across a Feshbach resonance, the BCS
pairs are projected onto Feshbach molecules, which can be further
probed for example by time-of-flight images \cite{liu:013607}. The bosons produced effectively reside in $p$ band and are stable, since by Pauli blocking the filled $s$-band fermions will prevent the the $p$-wave bosons from decaying \cite{liu:013607}. Here, we
use a simple model \cite{Diener2004,Altman2005} to evaluate
the momentum distribution of molecules after projection
\begin{equation}
n_{\mathbf{q}}=\sum_{\mathbf{k,k'}} f^*_{\mathbf{k}}f_{\mathbf{k'}}\langle
S^{\dag}_{\mathbf{k+q}/2}P^{\dag}_{\mathbf{-k+q}/2}P_{\mathbf{-k'+q}/2} S_{\mathbf{k'+q}/2}\rangle,
\label{eq:molecule}
\end{equation}
where $f_{\mathbf{k}}$ is the molecular wave function, and the correlation function can be evaluated
within mean field theory \cite{Altman2005}. For fast sweeps, the molecular size is small compared to
lattice constant and its wave function can be approximated by a delta function in real space (a
constant $\sqrt{1/N}$ in momentum space). By this assumption, $n_q$ is the same quantity as $C_q$ in
Eq.~\eqref{eq:moleculedmft}. Figure~\ref{fig:pair}(a) shows the $n_q$ of $p$-wave Feshbach molecules and
a peak is located at $0.433\pi/a$.  Figure~\ref{fig:pair}(b) shows $C_q$ from  Eq.~\eqref{eq:moleculedmft}, based on the DMRG results shown in Fig.~\ref{fig:cartoon}(b).The time-of-flight experiment is predicted to distribute peaks
corresponding to that in Fig.~\ref{fig:pair}. Note that for the 1D problem (Fig.~\ref{fig:pair}(b)),
the delta-function peak is replaced by a cusp characteristic of power law due to the lack of long range order.

We thank Chungwei Lin for helpful discussions. This work is supported by ARO Grant No.
W911NF-07-1-0293.

\bibliography{splusp}

\begin{thebibliography}{26}%
\makeatletter
\providecommand \@ifxundefined [1]{%
 \@ifx{#1\undefined}
}%
\providecommand \@ifnum [1]{%
 \ifnum #1\expandafter \@firstoftwo
 \else \expandafter \@secondoftwo
 \fi
}%
\providecommand \@ifx [1]{%
 \ifx #1\expandafter \@firstoftwo
 \else \expandafter \@secondoftwo
 \fi
}%
\providecommand \natexlab [1]{#1}%
\providecommand \enquote  [1]{``#1''}%
\providecommand \bibnamefont  [1]{#1}%
\providecommand \bibfnamefont [1]{#1}%
\providecommand \citenamefont [1]{#1}%
\providecommand \href@noop [0]{\@secondoftwo}%
\providecommand \href [0]{\begingroup \@sanitize@url \@href}%
\providecommand \@href[1]{\@@startlink{#1}\@@href}%
\providecommand \@@href[1]{\endgroup#1\@@endlink}%
\providecommand \@sanitize@url [0]{\catcode `\\12\catcode `\$12\catcode
  `\&12\catcode `\#12\catcode `\^12\catcode `\_12\catcode `\%12\relax}%
\providecommand \@@startlink[1]{}%
\providecommand \@@endlink[0]{}%
\providecommand \url  [0]{\begingroup\@sanitize@url \@url }%
\providecommand \@url [1]{\endgroup\@href {#1}{\urlprefix }}%
\providecommand \urlprefix  [0]{URL }%
\providecommand \Eprint [0]{\href }%
\providecommand \doibase [0]{http://dx.doi.org/}%
\providecommand \selectlanguage [0]{\@gobble}%
\providecommand \bibinfo  [0]{\@secondoftwo}%
\providecommand \bibfield  [0]{\@secondoftwo}%
\providecommand \translation [1]{[#1]}%
\providecommand \BibitemOpen [0]{}%
\providecommand \bibitemStop [0]{}%
\providecommand \bibitemNoStop [0]{.\EOS\space}%
\providecommand \EOS [0]{\spacefactor3000\relax}%
\providecommand \BibitemShut  [1]{\csname bibitem#1\endcsname}%
\let\auto@bib@innerbib\@empty
\bibitem [{\citenamefont {Casalbuoni}\ and\ \citenamefont
  {Nardulli}(2004)}]{RevModPhys.76.263}%
  \BibitemOpen
  \bibfield  {author} {\bibinfo {author} {\bibfnamefont {R.}~\bibnamefont
  {Casalbuoni}}\ and\ \bibinfo {author} {\bibfnamefont {G.}~\bibnamefont
  {Nardulli}},\ }\href@noop {} {\bibfield  {journal} {\bibinfo  {journal} {Rev.
  Mod. Phys.}\ }\textbf {\bibinfo {volume} {76}},\ \bibinfo {pages} {263}
  (\bibinfo {year} {2004})}\BibitemShut {NoStop}%
\bibitem [{\citenamefont {Giorgini}\ \emph {et~al.}(2008)\citenamefont
  {Giorgini}, \citenamefont {Pitaevskii},\ and\ \citenamefont
  {Stringari}}]{giorgini:1215}%
  \BibitemOpen
  \bibfield  {author} {\bibinfo {author} {\bibfnamefont {S.}~\bibnamefont
  {Giorgini}}, \bibinfo {author} {\bibfnamefont {L.~P.}\ \bibnamefont
  {Pitaevskii}}, \ and\ \bibinfo {author} {\bibfnamefont {S.}~\bibnamefont
  {Stringari}},\ }\href@noop {} {\bibfield  {journal} {\bibinfo  {journal}
  {Rev. Mod. Phys.}\ }\textbf {\bibinfo {volume} {80}},\ \bibinfo {eid} {1215}
  (\bibinfo {year} {2008})}\BibitemShut {NoStop}%
\bibitem [{\citenamefont {Sheehy}\ and\ \citenamefont
  {Radzihovsky}(2007)}]{Sheehy20071790}%
  \BibitemOpen
  \bibfield  {author} {\bibinfo {author} {\bibfnamefont {D.~E.}\ \bibnamefont
  {Sheehy}}\ and\ \bibinfo {author} {\bibfnamefont {L.}~\bibnamefont
  {Radzihovsky}},\ }\href@noop {} {\bibfield  {journal} {\bibinfo  {journal}
  {Ann. Phys. (NY)}\ }\textbf {\bibinfo {volume} {322}},\ \bibinfo {pages}
  {1790 } (\bibinfo {year} {2007})}\BibitemShut {NoStop}%
\bibitem [{\citenamefont {Radzihovsky}\ and\ \citenamefont
  {Sheehy}(2010)}]{0034-4885-73-7-076501}%
  \BibitemOpen
  \bibfield  {author} {\bibinfo {author} {\bibfnamefont {L.}~\bibnamefont
  {Radzihovsky}}\ and\ \bibinfo {author} {\bibfnamefont {D.~E.}\ \bibnamefont
  {Sheehy}},\ }\href@noop {} {\bibfield  {journal} {\bibinfo  {journal} {Rep.
  Prog. Phys.}\ }\textbf {\bibinfo {volume} {73}},\ \bibinfo {pages} {076501}
  (\bibinfo {year} {2010})}\BibitemShut {NoStop}%
\bibitem [{\citenamefont {Parish}\ \emph {et~al.}(2007)\citenamefont {Parish},
  \citenamefont {Baur}, \citenamefont {Mueller},\ and\ \citenamefont
  {Huse}}]{PhysRevLett.99.250403}%
  \BibitemOpen
  \bibfield  {author} {\bibinfo {author} {\bibfnamefont {M.~M.}\ \bibnamefont
  {Parish}}, \bibinfo {author} {\bibfnamefont {S.~K.}\ \bibnamefont {Baur}},
  \bibinfo {author} {\bibfnamefont {E.~J.}\ \bibnamefont {Mueller}}, \ and\
  \bibinfo {author} {\bibfnamefont {D.~A.}\ \bibnamefont {Huse}},\ }\href@noop
  {} {\bibfield  {journal} {\bibinfo  {journal} {Phys. Rev. Lett.}\ }\textbf
  {\bibinfo {volume} {99}},\ \bibinfo {pages} {250403} (\bibinfo {year}
  {2007})}\BibitemShut {NoStop}%
\bibitem [{\citenamefont {Fulde}\ and\ \citenamefont
  {Ferrell}(1964)}]{PhysRev.135.A550}%
  \BibitemOpen
  \bibfield  {author} {\bibinfo {author} {\bibfnamefont {P.}~\bibnamefont
  {Fulde}}\ and\ \bibinfo {author} {\bibfnamefont {R.~A.}\ \bibnamefont
  {Ferrell}},\ }\href@noop {} {\bibfield  {journal} {\bibinfo  {journal} {Phys.
  Rev.}\ }\textbf {\bibinfo {volume} {135}},\ \bibinfo {pages} {A550} (\bibinfo
  {year} {1964})}\BibitemShut {NoStop}%
\bibitem [{ffl()}]{fflo2}%
  \BibitemOpen
  \href@noop {} {}\bibinfo {note} {A. I. Larkin and Y. N. Ovchinnikov, Zh.
  Eksp. Teor. Fiz. {\bf 47}, 1136 (1964) [Sov. Phys. JETP {\bf 20}, 762
  (1965)].}\BibitemShut {Stop}%
\bibitem [{\citenamefont {M\"uther}\ and\ \citenamefont
  {Sedrakian}(2002)}]{PhysRevLett.88.252503}%
  \BibitemOpen
  \bibfield  {author} {\bibinfo {author} {\bibfnamefont {H.}~\bibnamefont
  {M\"uther}}\ and\ \bibinfo {author} {\bibfnamefont {A.}~\bibnamefont
  {Sedrakian}},\ }\href@noop {} {\bibfield  {journal} {\bibinfo  {journal}
  {Phys. Rev. Lett.}\ }\textbf {\bibinfo {volume} {88}},\ \bibinfo {pages}
  {252503} (\bibinfo {year} {2002})}\BibitemShut {NoStop}%
\bibitem [{\citenamefont {Sedrakian}\ \emph {et~al.}(2005)\citenamefont
  {Sedrakian}, \citenamefont {Mur-Petit}, \citenamefont {Polls},\ and\
  \citenamefont {M\"uther}}]{PhysRevA.72.013613}%
  \BibitemOpen
  \bibfield  {author} {\bibinfo {author} {\bibfnamefont {A.}~\bibnamefont
  {Sedrakian}}, \bibinfo {author} {\bibfnamefont {J.}~\bibnamefont
  {Mur-Petit}}, \bibinfo {author} {\bibfnamefont {A.}~\bibnamefont {Polls}}, \
  and\ \bibinfo {author} {\bibfnamefont {H.}~\bibnamefont {M\"uther}},\
  }\href@noop {} {\bibfield  {journal} {\bibinfo  {journal} {Phys. Rev. A}\
  }\textbf {\bibinfo {volume} {72}},\ \bibinfo {pages} {013613} (\bibinfo
  {year} {2005})}\BibitemShut {NoStop}%
\bibitem [{\citenamefont {Liu}\ and\ \citenamefont
  {Wilczek}(2003)}]{PhysRevLett.90.047002}%
  \BibitemOpen
  \bibfield  {author} {\bibinfo {author} {\bibfnamefont {W.~V.}\ \bibnamefont
  {Liu}}\ and\ \bibinfo {author} {\bibfnamefont {F.}~\bibnamefont {Wilczek}},\
  }\href@noop {} {\bibfield  {journal} {\bibinfo  {journal} {Phys. Rev. Lett.}\
  }\textbf {\bibinfo {volume} {90}},\ \bibinfo {pages} {047002} (\bibinfo
  {year} {2003})}\BibitemShut {NoStop}%
\bibitem [{\citenamefont {Forbes}\ \emph {et~al.}(2005)\citenamefont {Forbes},
  \citenamefont {Gubankova}, \citenamefont {Liu},\ and\ \citenamefont
  {Wilczek}}]{PhysRevLett.94.017001}%
  \BibitemOpen
  \bibfield  {author} {\bibinfo {author} {\bibfnamefont {M.~M.}\ \bibnamefont
  {Forbes}}, \bibinfo {author} {\bibfnamefont {E.}~\bibnamefont {Gubankova}},
  \bibinfo {author} {\bibfnamefont {W.~V.}\ \bibnamefont {Liu}}, \ and\
  \bibinfo {author} {\bibfnamefont {F.}~\bibnamefont {Wilczek}},\ }\href@noop
  {} {\bibfield  {journal} {\bibinfo  {journal} {Phys. Rev. Lett.}\ }\textbf
  {\bibinfo {volume} {94}},\ \bibinfo {pages} {017001} (\bibinfo {year}
  {2005})}\BibitemShut {NoStop}%
\bibitem [{\citenamefont {Schirotzek}\ \emph {et~al.}(2009)\citenamefont
  {Schirotzek}, \citenamefont {Wu}, \citenamefont {Sommer},\ and\ \citenamefont
  {Zwierlein}}]{PhysRevLett.102.230402}%
  \BibitemOpen
  \bibfield  {author} {\bibinfo {author} {\bibfnamefont {A.}~\bibnamefont
  {Schirotzek}}, \bibinfo {author} {\bibfnamefont {C.-H.}\ \bibnamefont {Wu}},
  \bibinfo {author} {\bibfnamefont {A.}~\bibnamefont {Sommer}}, \ and\ \bibinfo
  {author} {\bibfnamefont {M.~W.}\ \bibnamefont {Zwierlein}},\ }\href@noop {}
  {\bibfield  {journal} {\bibinfo  {journal} {Phys. Rev. Lett.}\ }\textbf
  {\bibinfo {volume} {102}},\ \bibinfo {pages} {230402} (\bibinfo {year}
  {2009})}\BibitemShut {NoStop}%
\bibitem [{\citenamefont {Liu}\ and\ \citenamefont {Wu}(2006)}]{liu:013607}%
  \BibitemOpen
  \bibfield  {author} {\bibinfo {author} {\bibfnamefont {W.~V.}\ \bibnamefont
  {Liu}}\ and\ \bibinfo {author} {\bibfnamefont {C.}~\bibnamefont {Wu}},\
  }\href@noop {} {\bibfield  {journal} {\bibinfo  {journal} {Phys. Rev. A}\
  }\textbf {\bibinfo {volume} {74}},\ \bibinfo {eid} {013607} (\bibinfo {year}
  {2006})}\BibitemShut {NoStop}%
\bibitem [{\citenamefont {Isacsson}\ and\ \citenamefont
  {Girvin}(2005)}]{PhysRevA.72.053604}%
  \BibitemOpen
  \bibfield  {author} {\bibinfo {author} {\bibfnamefont {A.}~\bibnamefont
  {Isacsson}}\ and\ \bibinfo {author} {\bibfnamefont {S.~M.}\ \bibnamefont
  {Girvin}},\ }\href@noop {} {\bibfield  {journal} {\bibinfo  {journal} {Phys.
  Rev. A}\ }\textbf {\bibinfo {volume} {72}},\ \bibinfo {pages} {053604}
  (\bibinfo {year} {2005})}\BibitemShut {NoStop}%
\bibitem [{\citenamefont {Kuklov}(2006)}]{kuklov:110405}%
  \BibitemOpen
  \bibfield  {author} {\bibinfo {author} {\bibfnamefont {A.~B.}\ \bibnamefont
  {Kuklov}},\ }\href@noop {} {\bibfield  {journal} {\bibinfo  {journal} {Phys.
  Rev. Lett.}\ }\textbf {\bibinfo {volume} {97}},\ \bibinfo {eid} {110405}
  (\bibinfo {year} {2006})}\BibitemShut {NoStop}%
\bibitem [{\citenamefont {Zhao}\ and\ \citenamefont
  {Liu}(2008{\natexlab{a}})}]{zhao:160403}%
  \BibitemOpen
  \bibfield  {author} {\bibinfo {author} {\bibfnamefont {E.}~\bibnamefont
  {Zhao}}\ and\ \bibinfo {author} {\bibfnamefont {W.~V.}\ \bibnamefont {Liu}},\
  }\href@noop {} {\bibfield  {journal} {\bibinfo  {journal} {Phys. Rev. Lett.}\
  }\textbf {\bibinfo {volume} {100}},\ \bibinfo {eid} {160403} (\bibinfo {year}
  {2008}{\natexlab{a}})}\BibitemShut {NoStop}%
\bibitem [{\citenamefont {Wu}(2008)}]{wu:200406}%
  \BibitemOpen
  \bibfield  {author} {\bibinfo {author} {\bibfnamefont {C.}~\bibnamefont
  {Wu}},\ }\href@noop {} {\bibfield  {journal} {\bibinfo  {journal} {Phys. Rev.
  Lett.}\ }\textbf {\bibinfo {volume} {100}},\ \bibinfo {eid} {200406}
  (\bibinfo {year} {2008})}\BibitemShut {NoStop}%
\bibitem [{\citenamefont {Nikoli\ifmmode~\acute{c}\else \'{c}\fi{}}\ \emph
  {et~al.}(2010)\citenamefont {Nikoli\ifmmode~\acute{c}\else \'{c}\fi{}},
  \citenamefont {Burkov},\ and\ \citenamefont
  {Paramekanti}}]{PhysRevB.81.012504}%
  \BibitemOpen
  \bibfield  {author} {\bibinfo {author} {\bibfnamefont {P.}~\bibnamefont
  {Nikoli\ifmmode~\acute{c}\else \'{c}\fi{}}}, \bibinfo {author} {\bibfnamefont
  {A.~A.}\ \bibnamefont {Burkov}}, \ and\ \bibinfo {author} {\bibfnamefont
  {A.}~\bibnamefont {Paramekanti}},\ }\href@noop {} {\bibfield  {journal}
  {\bibinfo  {journal} {Phys. Rev. B}\ }\textbf {\bibinfo {volume} {81}},\
  \bibinfo {pages} {012504} (\bibinfo {year} {2010})}\BibitemShut {NoStop}%
\bibitem [{\citenamefont {Wu}(2009)}]{congjun:modphyslett}%
  \BibitemOpen
  \bibfield  {author} {\bibinfo {author} {\bibfnamefont {C.}~\bibnamefont
  {Wu}},\ }\href@noop {} {\bibfield  {journal} {\bibinfo  {journal} {Mod. Phys.
  Lett}\ }\textbf {\bibinfo {volume} {23}},\ \bibinfo {pages} {1} (\bibinfo
  {year} {2009})}\BibitemShut {NoStop}%
\bibitem [{\citenamefont {Radzihovsky}\ and\ \citenamefont
  {Choi}(2009)}]{radzihovsky:095302}%
  \BibitemOpen
  \bibfield  {author} {\bibinfo {author} {\bibfnamefont {L.}~\bibnamefont
  {Radzihovsky}}\ and\ \bibinfo {author} {\bibfnamefont {S.}~\bibnamefont
  {Choi}},\ }\href@noop {} {\bibfield  {journal} {\bibinfo  {journal} {Phys.
  Rev. Lett.}\ }\textbf {\bibinfo {volume} {103}},\ \bibinfo {eid} {095302}
  (\bibinfo {year} {2009})}\BibitemShut {NoStop}%
\bibitem [{\citenamefont {Feiguin}\ and\ \citenamefont
  {Heidrich-Meisner}(2009)}]{feiguin:076403}%
  \BibitemOpen
  \bibfield  {author} {\bibinfo {author} {\bibfnamefont {A.~E.}\ \bibnamefont
  {Feiguin}}\ and\ \bibinfo {author} {\bibfnamefont {F.}~\bibnamefont
  {Heidrich-Meisner}},\ }\href@noop {} {\bibfield  {journal} {\bibinfo
  {journal} {Phys. Rev. Lett.}\ }\textbf {\bibinfo {volume} {102}},\ \bibinfo
  {eid} {076403} (\bibinfo {year} {2009})}\BibitemShut {NoStop}%
\bibitem [{foo()}]{footnote}%
  \BibitemOpen
  \href@noop {} {}\bibinfo {note} {We have tried various parameters in the DMRG
  and mean field calculations for 1D and quasi-1D case respectively, and
  consistently found the $p$-orbital pair condensate.}\BibitemShut {Stop}%
\bibitem [{\citenamefont {Feiguin}\ and\ \citenamefont
  {Heidrich-Meisner}(2007)}]{feiguin:220508}%
  \BibitemOpen
  \bibfield  {author} {\bibinfo {author} {\bibfnamefont {A.~E.}\ \bibnamefont
  {Feiguin}}\ and\ \bibinfo {author} {\bibfnamefont {F.}~\bibnamefont
  {Heidrich-Meisner}},\ }\href@noop {} {\bibfield  {journal} {\bibinfo
  {journal} {Phys. Rev. B}\ }\textbf {\bibinfo {volume} {76}},\ \bibinfo {eid}
  {220508} (\bibinfo {year} {2007})}\BibitemShut {NoStop}%
\bibitem [{\citenamefont {Zhao}\ and\ \citenamefont
  {Liu}(2008{\natexlab{b}})}]{zhao:063605}%
  \BibitemOpen
  \bibfield  {author} {\bibinfo {author} {\bibfnamefont {E.}~\bibnamefont
  {Zhao}}\ and\ \bibinfo {author} {\bibfnamefont {W.~V.}\ \bibnamefont {Liu}},\
  }\href@noop {} {\bibfield  {journal} {\bibinfo  {journal} {Phys. Rev. A}\
  }\textbf {\bibinfo {volume} {78}},\ \bibinfo {eid} {063605} (\bibinfo {year}
  {2008}{\natexlab{b}})}\BibitemShut {NoStop}%
\bibitem [{\citenamefont {Diener}\ and\ \citenamefont {Ho}(2004)}]{Diener2004}%
  \BibitemOpen
  \bibfield  {author} {\bibinfo {author} {\bibfnamefont {R.~B.}\ \bibnamefont
  {Diener}}\ and\ \bibinfo {author} {\bibfnamefont {T.-L.}\ \bibnamefont
  {Ho}},\ }\href@noop {} {\bibfield  {journal} {\bibinfo  {journal}
  {arXiv:cond-mat/0404517}\ } (\bibinfo {year} {2004})}\BibitemShut {NoStop}%
\bibitem [{\citenamefont {Altman}\ and\ \citenamefont
  {Vishwanath}(2005)}]{Altman2005}%
  \BibitemOpen
  \bibfield  {author} {\bibinfo {author} {\bibfnamefont {E.}~\bibnamefont
  {Altman}}\ and\ \bibinfo {author} {\bibfnamefont {A.}~\bibnamefont
  {Vishwanath}},\ }\href@noop {} {\bibfield  {journal} {\bibinfo  {journal}
  {Phys. Rev. Lett.}\ }\textbf {\bibinfo {volume} {95}},\ \bibinfo {pages}
  {110404} (\bibinfo {year} {2005})}\BibitemShut {NoStop}%
\end{thebibliography}%
\end{document}